\documentclass[preprint,prl,showpacs]{revtex4}
\usepackage{graphicx}
\usepackage{dcolumn}
\usepackage{bm}

\begin{document}
\preprint{APS/123-QED}
\author{P. Gegenwart$^{(1)}$, J. Custers$^{(1)}$, C. Geibel$^{(1)}$,
K. Neumaier$^{(2)}$, T. Tayama$^{(1,3)}$, K. Tenya$^{(1)}$, O.
Trovarelli$^{(1)}$, and F. Steglich$^{(1)}$}
\address{$^{(1)}$ Max-Planck Institute for Chemical Physics
of Solids, D-01187 Dresden, Germany
\\ $^{(2)}$Walther Meissner Institute, D-85748 Garching, Germany
\\ $^{(3)}$Institute of Solid State Physics, University of Tokyo,
Kashiwa, Chiba 277-8581, Japan}
\title{Magnetic-Field Induced Quantum Critical Point in YbRh$_2$Si$_2$}

\date{
\today%
}

\begin{abstract}
We report low-temperature calorimetric, magnetic and resistivity
measurements on the antiferromagnetic (AF) heavy-fermion metal
YbRh$_2$Si$_2$ (${T_N =}$ 70 mK) as a function of magnetic field $B$. While
for fields exceeding the critical value ${B_{c0}}$ at which
${T_N\rightarrow0}$ the low temperature resistivity shows an ${AT^2}$
dependence, a ${1/(B-B_{c0})}$ divergence of ${A(B)}$ upon reducing $B$ to
${B_{c0}}$ suggests singular scattering at the whole Fermi surface and a
divergence of the heavy quasiparticle mass. The observations are interpreted
in terms of a new type of quantum critical point separating a weakly AF
ordered from a weakly polarized heavy Landau-Fermi liquid state.
\end{abstract}

 \maketitle

The study of quantum phase transitions has attracted the interest of many
researchers in the last decades, especially since the discovery of the
cuprate superconductors. Quantum phase transitions, in contrast to their
classical counterparts at ${T > 0}$ where thermal fluctuations are
important, are driven by a control parameter other than temperature, e.g.,
composition or pressure. A quantum critical point (QCP) commonly separates
an ordered from a disordered phase at zero temperature. To study quantum
critical behavior the heavy-fermion (HF) systems are very suitable since
they can be tuned continuously from an antiferromagnetic (AF) to a
paramagnetic (PM) metallic state by the variation of a single parameter,
i.e., the strength of the ${4f}$-conduction electron hybridization $g$,
which can be modified by the application of either external pressure or
chemical substitution. According to itinerant spinfluctuation theory
\cite{Hertz,Millis,Moriya}, close to the critical value ${g_c}$ at which
${T_N \rightarrow 0}$, the abundance of low-lying and long-range spin
fluctuations, mediating the interactions between the heavy quasiparticles
(QP), gives rise to pronounced deviations from Landau Fermi liquid (LFL)
behavior. Instead of being constant as for a LFL, the QP mass and QP-QP
scattering cross section, being proportional to the low-temperature
coefficients of the electronic specific heat ${\gamma(T) = C_{el}(T)/T}$ and
the electrical resistivity ${A(T)=(\rho(T)-\rho_0)/T^2=\Delta\rho/T^2}$,
respectively, show a strong increase or even divergence upon cooling to
lowest temperatures. The origin of non-Fermi liquid (NFL) behavior, though
observed in an increasing number of HF systems \cite{Stewart}, is still a
subject of controversy \cite{Coleman}.

Another type of QCP arises by the suppression of AF order upon applying
magnetic fields $B$. By tuning ${T_N}$ towards zero temperature at a
critical field ${B_{c0}=B_c(0)}$, the AF correlations between the ordered
moments are suppressed resulting in a field-aligned (FA) state for ${B \geq
B_{c0}}$. This is very different to the destruction of the ordered moments
which occurs at ${B=0}$ upon "$g$-tuning" an antiferromagnet through its QCP
at ${g_c}$ as described above. Up to now, theoretical models for the QCP at
${B_{c0}}$ are lacking, and only the doped AF systems CeCu$_{6-x}$Ag$_x$
\cite{Heuser98a,Heuser98b} and YbCu$_{5-x}$Al$_x$ \cite{Seuring} had been
tuned by magnetic field through this kind of QCP. For the latter system a
substantial amount of Al-doping is necessary to induce long-range AF order
leading to a broad phase-transition anomaly in zero field. It is not clear
in this case, how the observed NFL behavior is influenced by disorder. For
single crystalline CeCu$_{5.2}$Ag$_{0.8}$, NFL effects were observed only
for fields applied along the easy magnetic direction and do not indicate a
divergence of either the QP mass or the QP-QP scattering cross-section
\cite{Heuser98b}. The results were described within the framework of the
self-consistent renormalization theory \cite{Moriya} originally developed
for the "$g$-tuned" QCP.

In this letter we concentrate on the HF metal YbRh$_2$Si$_2$ for which
pronounced NFL phenomena, i.e., a logarithmic increase of ${C_{el}(T)/T}$
and a quasi-linear $T$-dependence of the electrical resistivity below 10 K,
have been observed above a low-lying AF phase transition \cite{Trovarelli
Letter}. This system is highly suited to study the properties of a
$B$-induced QCP, because i) the AF phase transition is of second order (see
below) and the ordering temperature ${T_N = 70}$ mK is the lowest among all
undoped HF systems at ambient pressure, ii) already very small magnetic
fields are sufficient to suppress the AF state, and iii) clean single
crystals can be studied, showing very sharp and well defined
phase-transition anomalies which do not broaden significantly at finite
fields \cite{NMR}. The application of pressure to YbRh$_2$Si$_2$ increases
${T_N}$ \cite{Trovarelli Letter} as expected, because the ionic volume of
the magnetic ${4f^{13}}$ Yb$^{3+}$-configuration is smaller than that of the
nonmagnetic ${4f^{14}}$ Yb$^{2+}$ one. Expanding the crystal lattice by
randomly substituting Ge for the smaller isoelectric Si atoms allows one to
"$g$-tune" YbRh$_2$(Si$_{1-x}$Ge$_x$)$_2$ through the QCP at ${x_c = (0.06
\pm 0.01)}$ without affecting its electronic properties and, due to the low
value of ${x_c}$, without introducing significant disorder to the lattice
\cite{Trovarelli Physica}. In YbRh$_2$(Si$_{0.95}$Ge$_{0.05}$)$_2$ the NFL
behavior extends to the lowest accessible temperatures, in particular
${\Delta\rho(T) \sim T}$ is observed from above 10 K to below 10 mK
\cite{Trovarelli Physica}. In the following, we report on low-temperature
magnetic, thermodynamic and transport properties of undoped YbRh$_2$Si$_2$
which are used to characterize its field-induced QCP.

High-quality single crystalline platelets of YbRh$_2$Si$_2$ were
grown from In flux as described earlier \cite{Trovarelli Letter}.
The new generation of crystals show a residual resistivity
${\rho_0 \simeq 1 \mu\Omega}$cm, i.e., twice as low as ${\rho_0}$
of the previous ones. Whereas for the latter no phase transition
anomaly at ${T_N}$ could be resolved in the resistivity
\cite{Trovarelli Letter}, the new crystals show a clear kink of
${\rho(T)}$ at ${T_N}$, see below. For all low-temperature
measurements, $^3$He/$^4$He dilution refrigerators were used. The
electrical resistivity and magnetic AC-susceptibility
${\chi_{AC}}$ were measured utilizing a Linear Research Co.
(LR700) bridge at 16.67 Hz. Amplitudes of 0.1 mA and 1 Oe for the
current and magnetic field, respectively, were chosen to determine
$\rho$ and ${\chi_{AC}}$. The DC-magnetization, ${M_{DC}}$,
measurements were performed utilizing a high-resolution capacitive
Faraday magnetometer as described in \cite{Sakakibara}. The
specific heat was determined with the aid of a quasi-adiabatic
heat pulse technique.

YbRh$_2$Si$_2$ exhibits a highly anisotropic magnetic response, indicating
that Yb$^{3+}$ moments are forming an "easy-plane" square lattice
perpendicular to the crystallographic c-direction \cite{Trovarelli Letter}.
We first discuss the magnetic properties measured with the field applied
along the easy tetragonal plane, ${B \perp c}$. At ${B \simeq 0}$,
${\chi_{AC}(T)}$ reveals a sharp AF phase-transition at ${T_N =}$ 70 mK
(Fig. 1a). In the paramagnetic state at ${T_N \leq  T \leq 0.6}$ K, the
susceptibility follows a Curie-Weiss type behavior implying fluctuating
moments of the order of ${1.4 \mu_B / Yb^{3+}}$-ion and a Weiss-temperature
of ${\Theta \simeq -0.32}$ K. The isothermal magnetization (Fig. 1b) shows a
strongly nonlinear response for fields ${B \perp c}$. For ${T < T_N}$ a
clear reduction in slope is observed above 0.06 T which indicates the
suppression of AF order and the transition into the FA state. A smooth
extrapolation of ${M_{DC}(B)}$ for ${B > 0.06}$ T towards zero field reveals
a value of ${\mu_s < 0.1 \mu_B}$ for the staggered magnetization in the AF
state, indicating that the size of the ordered moments is much smaller than
that of the effective moments observed in the PM state above ${T_N}$. Thus a
large fraction of the local moments appears to remain fluctuating within the
easy plane in the AF ordered state. Their continuous polarization for fields
exceeding ${B_{c0}}$ gives rise to a strong curvature in ${M(B)}$ for ${B
\perp c}$. For fields applied along the magnetic hard direction, ${B
\parallel c}$, the magnetization shows an almost linear behavior (Fig. 1b)
which was found to extend at least up to 58 T \cite{Custers}. At
${T < T_N}$ a very tiny decrease in the ${M(B)}$ slope is observed
at about 0.7 T which, according to the resistivity measurements
discussed below, represents the critical field ${B_{c0}}$ for ${B
\parallel c}$.

The low-$T$ resistivity was measured in magnetic fields applied both
perpendicular and parallel to the $c$-direction, with the current
perpendicular to the field direction in each case (Fig. 2). For ${B = 0}$
the resistivity follows a quasi-linear $T$-dependence down to about 80 mK,
where a sharp decrease, independent of the current direction, is observed.
We note that this behavior is not consistent with that observed for SDW
systems for which an increase of ${\rho(T)}$ along the direction of the SDW
modulation, indicating the partial gapping of the Fermi surface, should be
expected. The absence of this behavior favors the interpretation of
local-moment type of magnetic order in YbRh$_2$Si$_2$, compatible with the
observation of large fluctuating moments in ${\chi_{AC}(T)}$ above ${T_N}$.
The resistivity in the AF ordered state is best described by ${\Delta\rho
=AT^2}$ with a huge coefficient, ${A=22}$ ${\mu \Omega cm}$/K$^2$, for 20 mK
${\leq T \leq 60}$ mK. With increasing $B$, the phase-transition anomaly in
${\rho(T)}$ shifts to lower temperatures and vanishes at critical fields
${B_{c0}}$ of about 0.06 T and 0.66 T for ${B\perp c}$ and ${B\parallel c}$,
respectively. At ${B = B_{c0}}$, the resistivity follows a linear
$T$-dependence down to the lowest accessible temperature of about 20 mK.
This observation provides striking evidence for field-induced NFL behavior
at magnetic fields applied along both crystallographic directions. At ${B >
B_{c0}}$, we find ${\Delta\rho = A(B)\cdot T^2}$ for ${T \leq T^\ast(B)}$,
with ${T^\ast (B)}$ increasing and ${A(B)}$ decreasing upon raising the
applied magnetic field.

Next we turn to our low-temperature specific-heat results, ${C(T)}$, which
contain electronic and hyperfine contributions, while the phonon part can be
safely ignored. We use ${\Delta C=C-C_Q}$, where the nuclear quadrupolar
term calculated from recent Moessbauer results \cite{Abd} amounts to about
5\% of ${C(T)}$ at 40 mK. As reported in \cite{Trovarelli Letter}, the
zero-field ratio ${\Delta C/T}$ is proportional to ${-\log T}$ in a wide
temperature window, 0.3 K ${\leq T \leq 10}$ K, below which an additional
(as yet unexplained) upturn occurs. The new measurements, which were
performed in small magnetic fields and down to lower temperatures when
compared to the previous ones, show a clear mean-field-type anomaly at ${T =
T_N}$ (Fig. 3). Specific heat, therefore, confirms a second-order phase
transition, as already concluded from our magnetization measurements.
Extrapolating ${\Delta C(T)/T}$ as ${T\rightarrow 0}$ to ${\gamma_0 = (1.7
\pm 0.2)}$ J/K$^2$mol reveals an entropy gain at the AF phase transition of
only about ${0.01R\cdot\ln2}$. This is in accordance with the small value of
the staggered moments and gives further evidence for the weakness of the AF
order in YbRh$_2$Si$_2$. The ratio of ${A/\gamma_0^2}$ taken from the
${B=0}$ data in the ordered state is close to that for a LFL \cite{KW},
i.e., one with very heavy quasiparticle masses. At small magnetic fields
applied along the $c$-direction the phase-transition anomaly shifts to lower
$T$ as observed in both AC-susceptibility and resistivity experiments.
However, due to the strong magnetic anisotropy, the sample plate used for
the specific-heat measurement could not be aligned perfectly along the hard
magnetic direction. Therefore, a critical field ${B_{c0}}$ of only about 0.3
T was sufficient to suppress AF order completely in these experiments. As
shown in Fig. 3, at ${B = B_{c0}}$ the specific-heat coefficient ${\Delta
C(T)/T}$ increases down to the lowest $T$, indicative of a field-induced NFL
ground state. Within 40 mK ${\leq T \leq 120}$ mK it follows a steep
increase with a much larger slope than observed in zero-field at elevated
temperatures (see dotted line in Fig. 3 of Ref. \cite{Trovarelli Letter}).
While this anomalous contribution is strongly reduced upon increasing $B$,
at magnetic fields ${B \geq}$ 1 T, the nuclear contribution becomes visible
at the lowest temperatures, above which a constant ${\gamma_0(B)}$ value is
observed (Fig. 3). ${\gamma_0(B)}$ decreases in magnitude upon increasing
the field.

The results of the low-$T$ experiments are summarized in the ${T-B}$ phase
diagram displayed in Fig. 4. Here, the field dependence of the N\'{e}el
temperature was determined from the maximum value of the corresponding
d${\rho/}$d$T$ vs $T$ curves. For fields aligned in the easy plane, the
results agree perfectly well with those obtained from the AC-susceptibility
(Fig. 1a). For ${B > B_{c0}}$, the characteristic temperature ${T^\ast(B)}$
marks the upper limit of the observed ${T^2}$ behavior in the resistivity.

To study the nature of the field-induced QCP in YbRh$_2$Si$_2$, we analyze
the magnetic field dependence of the coefficients $A$, ${\gamma_0}$ and
${\chi_0}$ observed for ${T \rightarrow 0}$ in the resistivity, ${\Delta\rho
= A(B)T^2}$, specific heat, ${C/T=\gamma_0(B)}$ \cite{Trovarelli Letter},
and magnetic AC-susceptibility, ${\chi_{AC}=\chi_0(B)}$ \cite{Trovarelli
Letter} when approaching the QCP from the field-polarized state. As shown in
the inset of Fig. 5, we observe ${A\sim \gamma_0^2, A \sim \chi_0^2}$ and
thus also ${\gamma_0 \sim \chi_0}$, independent of the field orientation and
for all $B$ values exceeding ${B_{c0}}$. Thus, the FA state can be described
by the LFL model, too. Like in the AF ordered state (at ${B=0}$) we find
that the ${A/\gamma_0^2}$ ratio roughly equals that observed for many HF
systems \cite{KW}. This is in striking variance to the SDW scenario for a 2D
spin fluid and a 3D Fermi surface \cite{2dspinfluid}. Further on, a very
large Sommerfeld-Wilson ratio ${R =
(\chi_0/\gamma_0)(\pi^2k_B^2/\mu_0\mu_{eff}^2)}$ of about 14 ${(\mu_{eff} =
1.4 \mu_B)}$ indicates a strongly enhanced susceptibility in the
field-aligned state of YbRh$_2$Si$_2$ pointing to the importance of
low-lying ferromagnetic (${\bf q=0}$) fluctuations in YbRh$_2$Si$_2$
\cite{FM}. Since YbRh$_2$Si$_2$ behaves as a true LFL for ${B
> B_{c0}}$ and ${T < T^\ast(B)}$, the observed temperature dependences
should hold down to ${T = 0}$. The field dependence ${A(B)}$ shown
in Fig. 5 measures the QP-QP scattering cross section when, by
field tuning, crossing the QCP at zero temperature. Most
importantly, a ${1/(B-B_{c0})}$ divergence is observed indicating
that the whole Fermi surface undergoes singular scattering at the
$B$-tuned QCP. Furthermore, the relation ${A \sim \gamma_0^2}$
observed at elevated fields, suggests that also the QP mass
diverges, i.e., as ${1/(B-B_{c0})^{1/2}}$, when approaching
$B_{c0}$ \cite{CeB6}.

The extremely low value of the critical field applied along the easy plane
highlights the near degeneracy of two different LFL states, one being weakly
AF ordered (${B<B_{c0}}$) and the other one being weakly polarized
(${B>B_{c0}}$). In fact, the AC-susceptibility measured for fields near
${B_{c0}}$ shows a sharp increase when approaching the phase transition upon
cooling, see 0.04T data in Fig. 1a.

To conclude, a new type of QCP separating an antiferromagnetic from a
field-aligned ground state was studied in the clean heavy-fermion metal
YbRh$_2$Si$_2$. We observed pronounced NFL behavior at the critical field
${B_{c0}}$ necessary to suppress antiferromagnetic order. When this system
is tuned, at zero temperature, from the field-aligned state towards the QCP
by decreasing the applied magnetic field, both the quasiparticle scattering
cross section and the quasiparticle mass appear to diverge.

We are grateful to Piers Coleman, Qimiao Si, Greg Stewart and Heribert
Wilhelm for valuable discussions.

\newpage
\begin{figure}
\centerline{\includegraphics[width=.7\textwidth]{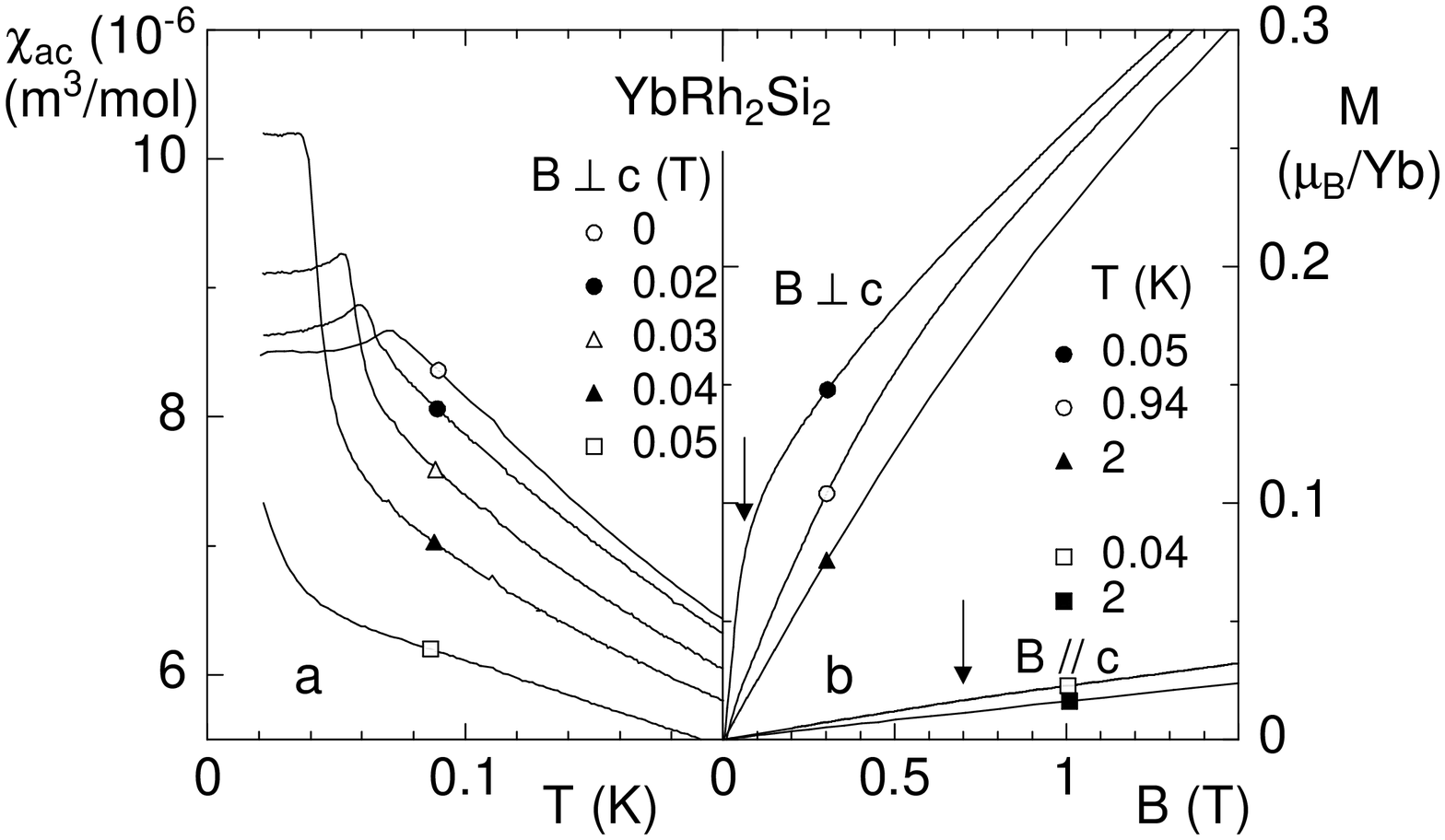}}
\caption{Low-temperature AC-susceptibility ${\chi_{AC}}$ of
YbRh$_2$Si$_2$ at varying fields applied perpendicular to the
$c$-axis (a) and isothermal DC magnetization ${M_{DC}}$ at varying
temperatures in magnetic fields applied along and perpendicular to
the $c$-axis. Arrows indicate critical fields ${B_{c0}}$.}
\label{fig1}
\end{figure}

\newpage
\begin{figure}
\centerline{\includegraphics[width=.7\textwidth]{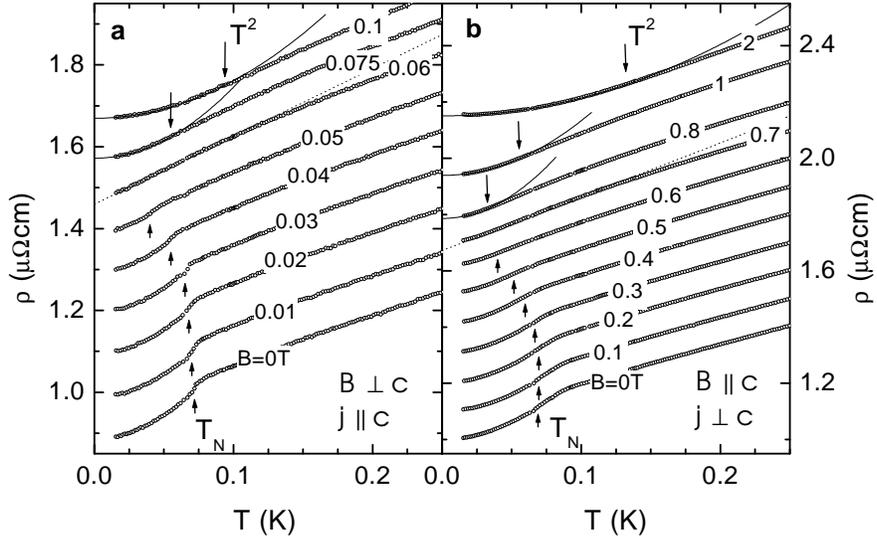}}
\caption{Low-temperature electrical resistivity of YbRh$_2$Si$_2$
at varying magnetic fields applied along the $a$- (a) and $c$-axis
(b). For clarity the different curves in ${B > 0}$ were shifted
subsequently by 0.1 ${\mu\Omega}$cm. Up- and downraising arrows
indicate ${T_N}$ - and upper limit of ${T^2}$ behavior,
respectively. Dotted and solid lines represent ${\Delta\rho\sim
T^\epsilon}$ with ${\epsilon=1}$ and ${\epsilon=2}$,
respectively.} \label{fig2}
\end{figure}

\newpage
\begin{figure}
\centerline{\includegraphics[width=.7\textwidth]{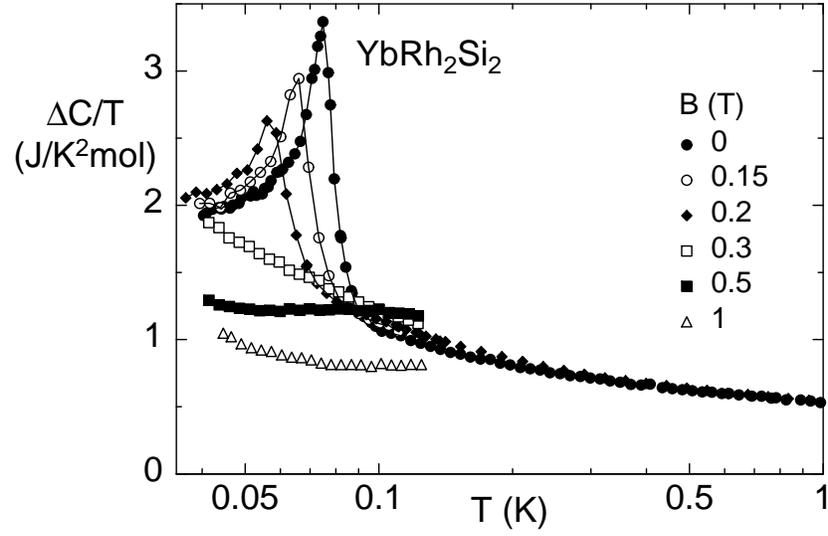}}
\caption{Specific heat as ${\Delta C/T=(C-C_Q)/T}$ vs $T$ (on a
logarithmic scale) for YbRh$_2$Si$_2$ at varying fields applied
parallel to the $c$-axis. ${C_Q \sim T^{-2}}$ is the nuclear
quadrupole contribution calculated from recent Moessbauer
results\cite{Abd}.} \label{fig3}
\end{figure}

\newpage
\begin{figure}
\centerline{\includegraphics[width=.7\textwidth]{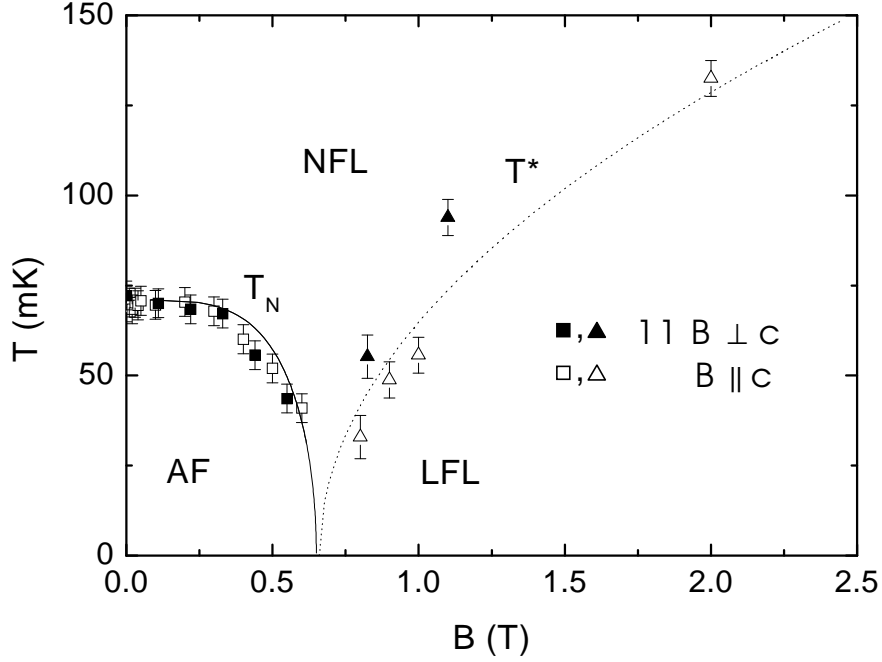}}
\caption{${T-B}$ phase diagram for YbRh$_2$Si$_2$ with ${T_N}$ as
derived from ${d\rho/dT}$ vs $T$ and ${T^\ast}$, the  upper limit
of the ${\Delta\rho = AT^2}$ behavior, as a function of magnetic
field, applied both along and perpendicular to the $c$-axis. For
the latter ones the $B$-values have been multiplied by a factor
11. Lines separating the antiferromagnetic (AF), non-Fermi liquid
(NFL) and Landau Fermi liquid (LFL) phase are guides to the eye.
Note that the AF phase transition as a function of field is a
continuous one, cf. ${M_{DC}(B)}$ curves in Fig. 1b.} \label{fig4}
\end{figure}

\newpage
\begin{figure}
\centerline{\includegraphics[width=.7\textwidth]{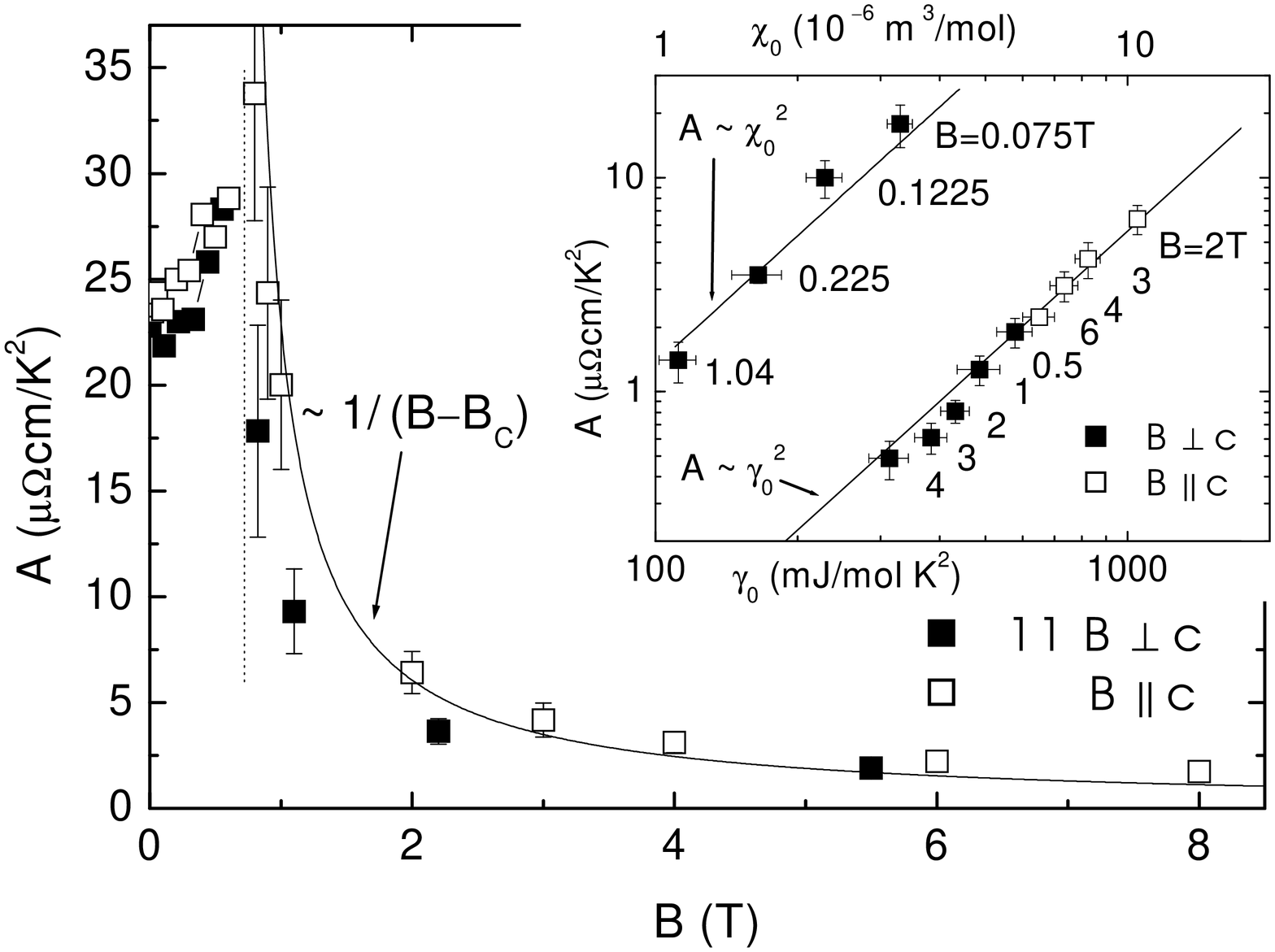}}
\caption{Coefficient ${A = \Delta\rho/T^2}$ vs field $B$. Data for
$B$ perpendicular to the $c$-direction have been multiplied by 11.
Dashed line marks ${B_{c0}}$, solid line represents
${(B-B_{c0})^{-1}}$. Inset shows double-$\log$ plot of $A$ vs
${\gamma_0}$ and $A$ vs ${\chi_0}$ for different magnetic fields.
Solid lines represent ${A/\gamma_0^2 = 5.8\cdot10^{-6}
\mu\Omega}$cm(Kmol/mJ)$^2$ and ${A/\chi_0^2 = 1.25\cdot10^{12}
\mu\Omega}$cmK$^{-2}$/(m$^3$/mol)$^2$.} \label{fig5}
\end{figure}


\begin{references}
\bibitem{Hertz} J.A. Hertz, Phs. Rev. {\bf B 14}, 1165 (1976).
\bibitem{Millis} A.J. Millis, Phys. Rev.{\bf B 48}, 7183 (1993).
\bibitem{Moriya} T. Moriya and T. Takimoto, J. Phys. Soc. Jpn. {\bf 64}, 960 (1995).
\bibitem{Stewart} G.R. Stewart, Rev. Mod. Phys. {\bf 73}, 797 (2001).
\bibitem{Coleman} P. Coleman et al., J. Phys. Cond. Matt. {\bf 13}, R723 (2001).
\bibitem{Heuser98a} K. Heuser et al., Phys. Rev. {\bf B 57}, R4198 (1998).
\bibitem{Heuser98b} K. Heuser et al., Phys. Rev. {\bf B 58}, R15959 (1998).
\bibitem{Seuring} S. Seuring et al., Physica {\bf B 281 \& 282}, 374 (2000).
\bibitem{Trovarelli Letter} O. Trovarelli et al., Phys. Rev. Lett. {\bf 85}, 626 (2000).
\bibitem{Trovarelli Physica} O. Trovarelli et al., Physica {\bf B 312-313}, 401 (2002).
\bibitem{NMR} A $^{29}$Si-NMR study of the field-tuned QCP has already been performed by Ishida et al. \cite{Ishida}.
This study is, however, restricted to ${B \geq}$ 0.15 T for fields applied
in the magnetic easy plane (${B\perp c}$).
\bibitem{Ishida} K. Ishida et al., to be published.
\bibitem{Sakakibara} T. Sakakibara et al., Jpn. J. Appl. Phys. {\bf 33}, 5067 (1994).
\bibitem{Custers} J. Custers et al., Act. Phys. Pol. {\bf B 32}, 3221 (2001).
\bibitem{Abd} M. Abd-Elmeguid (unpublished results).
\bibitem{KW} K. Kadowaki and S.B. Woods, Sol. St. Com. {\bf 58}, 507 (1986).
\bibitem{2dspinfluid} In this model the parameter $\delta$ which is the
square of the inverse magnetic correlation length in the 2D spin fluid
measures the distance from the QCP, i.e., ${\delta\sim(B-B_{c0})}$. Assuming
that the spin fluid renders the entire Fermi surface "hot", the coefficient
$A$ diverges as ${A\sim1/\delta}$, whereas for the specific heat coefficient
${\gamma_0}$ a much weaker divergence ${\gamma_0 \sim\ log(1/\delta)}$ is
expected \cite{Kotliar}. Thus, this model would predict the ratio
${A/\gamma_0^2}$ to diverge instead of being constant for ${B \rightarrow
B_{c0}}$.
\bibitem{Kotliar} I. Paul and G. Kotliar, Phys. Rev. {\bf B 64}, 184414
(2001).
\bibitem{FM} As also inferred from recent $^{29}$Si-NMR measurements \cite{Ishida}.
\bibitem{CeB6} Reminiscent of the mass divergence observed in
specific heat and dHvA experiments at the phase transition from AF
to antiferroquadrupolar order in CeB$_6$ at 2 T \cite{Joss}.
\bibitem{Joss} W. Joss et al., Phys. Rev. Lett. {\bf 59}, 1609
(1987).
\end{references}
\end{document}